\documentclass[final,11pt,twocolumn,sort&compress]{elsarticle} %ELSEVIER

\usepackage{listings}
\usepackage{amsmath}
\usepackage{graphicx}
\usepackage{color}
\usepackage{setspace}
\usepackage{url}
\usepackage{widetext} %ELSEVIER
\usepackage{alltt, parskip, fancyhdr, boxedminipage}
\usepackage{makeidx, multirow, longtable, tocbibind, amssymb}
\usepackage{hyperref} \hypersetup{colorlinks=true,citecolor=blue,linkcolor=blue,urlcolor=blue} % HYPERLINKS

\usepackage{relsize} % for AFLOWpi
\newcommand{\AFLOWpi}{ {\sf AFLOW$\mathlarger{\mathlarger{\mathlarger{{\pi}}}}$}}

\newlength{\funcindent}
\newlength{\funcwidth}
\DeclareFixedFont{\ttb}{T1}{txtt}{bx}{n}{12} % for bold
\DeclareFixedFont{\ttm}{T1}{txtt}{m}{n}{12}  % for normal

\usepackage{color}
\definecolor{deepblue}{rgb}{0,0,0.5}
\definecolor{deepred}{rgb}{0.6,0,0}
\definecolor{deepgreen}{rgb}{0,0.5,0}

\def\AFLOW{{\small AFLOW}}   % AFLOW-TEX COMMON
\def\citeAFLOW{\cite{aflowPAPER,aflowBZ,curtarolo:art110,curtarolo:art63,curtarolo:art57,curtarolo:art49,monsterPGM,curtarolo:art94}}      % AFLOW-TEX COMMON
\def\citeAFLOWLIB{\cite{aflowlibPAPER,curtarolo:art92,curtarolo:art104}} %,curtarolo:art124}}
\newcommand\pythonstyle{\lstset{
basicstyle=\ttb\scriptsize,	
keywordstyle=\ttb\scriptsize\color{deepblue},
commentstyle=\ttb\scriptsize\color{deepred},    % Custom highlighting style
stringstyle=\ttb\scriptsize\color{deepgreen},
 language=Python,
  aboveskip=0mm,
  belowskip=0mm,
  showstringspaces=false,
  columns=flexible,
  numbers=none,
  numberstyle=\tiny\color{gray},
  ndkeywords={return, class, if ,elif, endif, while, do, else, True, False , catch, def},
  ndkeywordstyle=\color{red}\bfseries,
  breaklines=true,
  postbreak=\raisebox{0ex}[0ex][0ex]{\ensuremath{\color{black}\hookrightarrow\space}},
  breakatwhitespace=True,
  tabsize=2
}}

\lstnewenvironment{python}[1][]
{
\pythonstyle
\lstset{#1}
}
{}
\newcommand\pythoninline[1]{{\pythonstyle\lstinline!#1!}}

\newcommand\fortranstyle{\lstset{
language=Fortran,
basicstyle=\ttb\scriptsize,
otherkeywords={self},             % Add keywords here
keywordstyle=\ttb\scriptsize\color{deepblue},
emph={MyClass,__init__},          % Custom highlighting
commentstyle=\ttb\scriptsize\color{deepred},    % Custom highlighting style
stringstyle=\ttb\scriptsize\color{deepgreen},
showstringspaces=false,
}}

\lstnewenvironment{fortran}[1][]
{
\fortranstyle
\lstset{#1}
}
{}
\newcommand\fortraninline[1]{{\fortranstyle\lstinline!#1!}}

\definecolor{Gray}{gray}{0.9}
\definecolor{LightCyan}{rgb}{0.88,1,1}
\begin{document} 
\begin{frontmatter} %ELSEVIER


%\titlespacing{\section}{0pt}{*0}{*0}
%\titlespacing{\subsection}{0pt}{*0}{*0}
%\titlespacing{\subsubsection}{0pt}{*0}{*0}

\title{\LARGE {\bf  \AFLOWpi: A minimalist approach to high-throughput {\it ab initio} calculations including the generation of tight-binding hamiltonians}
}

\author{Andrew R. Supka$^{1,2}$, Troy E. Lyons$^{1}$, Laalitha Liyanage$^{3}$, Pino D'Amico$^{4,5}$, Rabih Al Rahal Al Orabi$^{1}$, Sharad Mahatara$^{1}$, Priya Gopal$^{1}$, Cormac Toher$^{7,8}$, Davide Ceresoli$^{6}$, Arrigo Calzolari$^{3,4,5,8}$, Stefano Curtarolo$^{3,8}$, Marco Buongiorno Nardelli$^{2,8}$, and Marco Fornari$^{1,2,8\star}$}

\address{$^1$ Department of Physics, Central Michigan University, Mount Pleasant MI, USA}
\address{$^2$ Science of Advanced Materials Program, Central Michigan University, Mount Pleasant MI, USA}

\address{$^{3}$Department of Physics and Department of Chemistry, University of North Texas, Denton TX, USA }
\address{$^{4}$Dipartimento di Fisica, Informatica e Matematica, Universit\'a di Modena and Reggio Emilia, Via Campi 213/a, 41125 Modena, Italy}
\address{$^{5}$CNR-NANO Research Center S3, Via Campi 213/a, 41125 Modena, Italy}

\address{$^{6}$CNR-ISTM, Istituto di Scienze e Tecnologie Molecolari, I-20133 Milano, Italy}
\address{$^{7}$Materials Science, Electrical Engineering, Physics and Chemistry, Duke University, Durham NC, 27708}
\address{$^{8}$Center for Materials Genomics, Duke University, Durham, NC 27708, USA}
 
\address{$^{\star}${\bf corresponding:} marco.fornari@cmich.edu}
\begin{abstract}
Tight-binding models provide a conceptually transparent and computationally efficient method to represent the electronic properties of materials. With \AFLOWpi\  we introduce a framework for high-throughput first principles calculations that automatically generates tight-binding hamiltonians without any additional input.  Several additional features are included in \AFLOWpi\ with the intent to simplify the self-consistent calculation of Hubbard $U$ corrections, the calculations of phonon dispersions, elastic properties, complex dielectric constants, and electronic transport coefficients. As examples we show how to compute the optical properties of layered nitrides in the $AM$N$_2$ family, and the elastic and vibrational properties of binary halides with CsCl and NaCl structure.

  \end{abstract}

%\date{\today} %APS\maketitle %APS

 %ELSEVIER
 \end{frontmatter} 

\section{Introduction}
\label{introduction}
 
Accelerated materials discovery is currently the focus of much infrastructure development both for experimental (see for instance Refs. \citep{10.1021co200007w,doi:10.1146/annurev.matsci.38.060407.130217,takeuchi2005data,cite-key}) and theoretical research \cite{nmatHT,0957-0233-16-1-039,monster}. The key components always involve robust data generation or collection, real time feedback and error control, curation and archival of the data, and post-processing tools for analysis and visualization. The ultimate goal is to discover relationships that may be hidden in the large data sets and possibly predict materials with improved specific functionalities.

\begin{figure}
  \centering
  \includegraphics[width=0.49\textwidth]{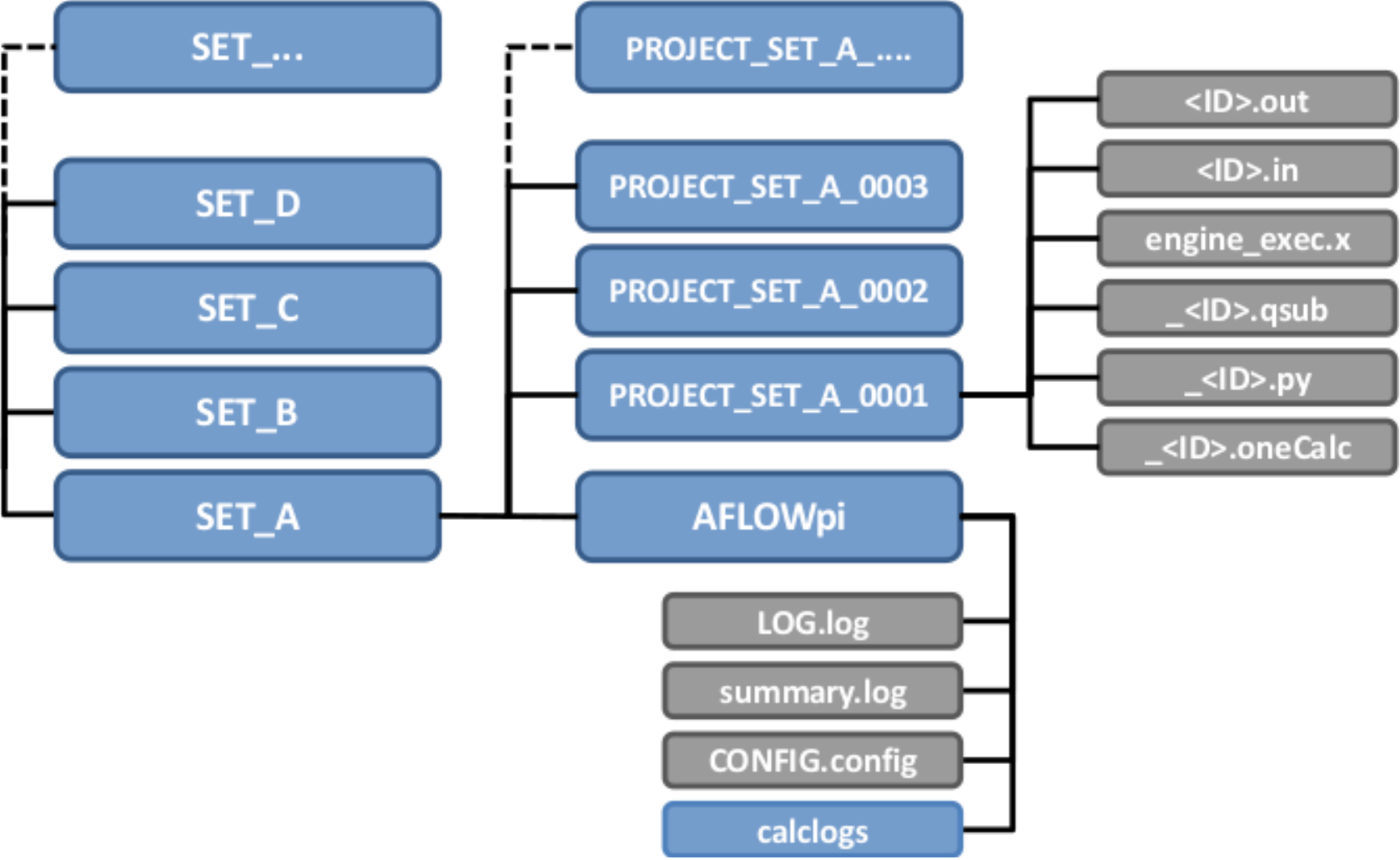}
  \caption{\small 
    The directory tree created by  \AFLOWpi\ during the preparation of the calculation with in a  {\tt PROJECT} directory. Each {\tt SET} contains the {\tt AFLOWpi} directory with several auxiliary files discussed in the text. {\tt PROJECT} and {\tt SET} are substituted with specific values.
  }
  \label{DIR_TREE}
\end{figure}

Flavors of density functional theory ({\small DFT}) usually form the basis for the generation of theoretical data since they provide reasonable transferability across different chemical compositions and structural themes at acceptable computational cost. Standard electronic structure packages such as Quantum Espresso \cite{qe} or {\small VASP} \cite{kresse_vasp_1} are used as the engine of the calculations but an additional layer of software tools is needed to efficiently manage the high-throughput (HT) workflows. In addition, to streamline data generation, those software tools usually curate the data and prepare them for distribution in online electronic structure properties databases \cite{materialsproject,aflowlibPAPER,nomad,acs.jctc.6b00114} 
%\cite{http://nomad-repository.eu/cms/} 
that can be mined {\it a posteriori}.

Several HT frameworks for electronic structure calculations were discussed in the literature.  
% in brief strength (and weakness) of these three tools mentioning and discussing the correspondent database 
{\small PYMATGEN} \cite{pymatgen} is the data generation software infrastructure behind the Materials Project database \cite{materialsproject}.
It is written in Python 2.7 using libraries such as Numpy and Scipy and exploits {\small VASP} and AbInit \cite{Gonze2002478} as standard electronic structure engines.
Many tools are available to study phase stability and phase diagrams.
% we used pythn 2.7 for library, numpy+scipy, pycifrw, bandsplot, rest api, interfaced with materialsprojectwebsite, self-healing
%what do they provide online?
% we don't do phase stability and phase diagrams
% weakness: VASP, 
AiiDA \cite{aiida} is an integrated software based on a paradigm involving automation, data, environment, and storage. It is written in Python 2.7 and revolves around relational databases for the overall design in addition to the storage component. For the automation component AiiDA involves workflows that use Quantum Espresso although other electronic structure codes are supported.  In addition, scripting interfaces such ASE \cite{ase} can be used by AiiDA to control the workflow.
% we used encapsulation of legacy code, workflows creation in Python, command line interface (CLI)
%  I didn't find a DB, http://www.aiida.net
%what do they provide online: nothing
%ADES: Automation, Data, Environment, and Sharing
% we don't want tight coupling of automation and storage, we don't use a daemon. This concept of integration with the DB is really pushed
% several schedulers are implemented SLURP, PBS
%very DB centered, integrated with SQL engines
%uses workflows
%querytools, several codes though ASE
ASE is paired with the Computational Materials Repository \cite{aserep}. 
Additional HT software tools for density functional calculations are  {\tt qmpy} (associated with the Open Quantum Materials Database) \cite{oqmd} and the HT toolkit (HTTK associated with the Open Materials Database, {\tt http://httk.opendatabase.se}); these software packages are written in Python and streamline electronic structure data generation.
%{\color{red} Add ASE-database, OQMD (Wolverton), HTTK (Rickard, Open Materials Database), Cambridge Materials Database}
The framework \AFLOW\ \citeAFLOW\ is an efficient tool for high-throughput calculations with VASP. It is written in C++ and includes tools to create input files for Quantum Espresso, to deal automatically with supercells and fractional occupations, to perform symmetry analysis, and to evaluate several thermodynamical and thermal transport properties at different levels of approximation \cite{curtarolo:art114,curtarolo:art96}. Data generated with \AFLOW\ are available online on the \AFLOW.org repositories \citeAFLOWLIB\ (see {\tt http://www.aflow.org}).

With \AFLOWpi\ we are complementing \AFLOW\ with a minimalist software infrastructure  that is easily portable, simple to use, and integrated with the \AFLOW.org repositories.
\AFLOWpi\ has been initially developed for verification and testing purposes but has evolved into a modular software infrastucture that  provides automatic workflows for the {\it ab initio} generation of tight-binding hamiltonians within the projected atomic orbital (PAO) scheme \cite{curtarolo:art86, curtarolo:art108, curtarolo:art111} and the self-consistent calculation of Hubbard $U$ corrections within the ACBN0 approach    
 \cite{curtarolo:art93}. In addition, workflows for the calculation of phonon dispersions with Hubbard $U$ correction, the calculation of elastic constants, diffusive transport coefficients, and optical spectra  are available. When possible, the code exploits the tight-binding hamiltonians as in Ref. \cite{pino}.

%In addition there are simple tools to compute descriptors \cite{naturecurtarolo} for thermal transport based on the quasi-harmonic approximations and the Debye-Callaway model \cite{debyecallaway1, debyecallaway2, rabih1,jose}.
% reference file centered, regular expression/text manipulation
% trivial to install, minimilist
% textbased, python module

\section{Software design and data organization}
\label{design}

\AFLOWpi 's design provides a simple, task oriented Python package to construct large sets of calculations from one or few  input files for a specific \textit{ab initio} engine and then to perform a workflow of tasks for each calculation in the set. 
The infrastructure is modeled on a simple and common {\it modus operandi} that involves prototypical input files (reference files) that are modified for desired tasks. The chosen workflow instructs \AFLOWpi\  how to modify the reference files and guides the process of submitting, running, and monitoring all calculations.  

The software is implemented in Python 2.7 (using the Python standard library and Numpy, Scipy, CIFfile, and Matplotlib). The systematic use of regular expression ({\tt re} module)  to parse and modify the input and output files according the specific workflow makes the software very portable and expandable to a variety of electronic structure engines and tools. Currently, \AFLOWpi\  encapsulates {\tt pw.x} and several post-processing software from the Quantum Espresso package \cite{qe} as well as {\tt ElaStic} \cite{elastic}. In addition it uses features of {\tt findsym} \cite{Stokes:zm5027}. The generation and manipulation of the TB Hamiltonians is done with the engine PAO$\pi$ \cite{paopi}.

The calculations and the resulting data are organized according to the layer structure adopted by \AFLOW\ \cite{curtarolo:art92}. 
The first layer in the directory hierarchy is the 
{\tt Project} layer that can contain multiple instances of the next layer, the 
{\tt Set} layer. 
The directories for the individual 
calculation pipelines in the set are named {\tt PROJECT\_SET\_XXXX} with 
{\tt PROJECT} being the project name, {\tt SET} being the name of the set and 
XXXX being the index of the 
calculation %. The indices of the calculations within a set are assigned as the calculations of a set are generated with the first generated being 0001, second  0002, etc. When no set is specified the form of the calculation directory name  is {\tt PROJECT__XXXX}  
(See Figure \ref{DIR_TREE}).
Each calculation pipeline receives a unique identifier to keep track of each pipe of calculations during 
setup and runtime as well as for storage and retrieval of the results. 
The identifier is constructed as a CRC64 hash \cite{CRC64}, which is a unique 16 character string representing the initial \textit{ab initio} engine input file for each calculation pipeline. The 
input file is first cleaned of commented out content and unnecessary whitespace, 
then lines of text are sorted alphabetically. Cleaning and sorting is necessary 
to ensure that extra whitespace or the ordering of the parameters will not 
change the value of the hash. 
Files for each task, or step, in each calculation pipeline are designated by {\tt ID\_XX} where 
{\tt XX} is the step number and ID is the identifier for that calculation 
pipeline. An input for the first step in a pipeline with the identifier 
{\tt 20b8c6f83a2ca9b} would read {\tt c20b8c6f83a2ca9b\_01.in}. 
Files and directories within each calculation's directory that are considered unneeded after runtime, often referred to as {\tt scratch}, such as Python scripts generated by \AFLOWpi, cluster submission 
files, and temporary files of the \textit{ab initio} engine have a prepended 
underscore. For example, the filename of the Python executable for the second 
step in that pipeline would be {\tt \_c20b8c6f83a2ca9b\_02.py}.
 \AFLOWpi\ modularizes the calculation pipelines by separating each task into its own executable. When a task is completed, the next task is initiated and so on until the pipeline completes. The location of the directory for each calculation pipeline is preserved for all tasks the user issues for the set. All files needed to perform the tasks as well as all files generated by the tasks are stored in the folder pertaining to the individual calculation pipeline in order to ensure control and transferability.		

Within each {\tt Set},  \AFLOWpi\ creates
a directory named {\tt AFLOWpi} 
that contains a master runtime log of the session, a copy of the configuration file used in the session, a summary log 
with relevant information from the output of every calculation within the set, 
as well as the loadable calculation files (LCFs).
In order to ensure data provenance {\tt AFLOWpi} includes mandatory information to be provided at the beginning of the data generation process. Each calculation pipeline has its own LCF that are pickled and used to control and monitor the workflow especially for resuming or expanding the calculations.
\AFLOWpi\  will run as a cluster job submission (currently PBS, {\tt http://pbsworks.com}) or as a local process on systems without a cluster queue. 
The user has the freedom to choose how the steps in the workflows of the calculations are grouped in terms of cluster submissions.

\begin{figure}[b!]
\begin{python}
import AFLOWpi
session = AFLOWpi.prep.init('mHT_TB', 'III-V',config='./mHT_TB.config')
allvars={}
allvars.update(
_AFLOWPI_A_ = ('Al','Ga','In'),
_AFLOWPI_B_ = ('P','As','Sb'),)
calcs = session.scfs(allvars,'mHT_TB.ref')
calcs.vcrelax()
tb_ham = calcs.tight_binding()
tb_ham.dos()
tb_ham.bands()
tb_ham.plot.bands(DOSPlot='APDOS')
calcs.acbn0()
tb_ham = calcs.tight_binding()
tb_ham.dos()
tb_ham.bands()
tb_ham.plot.bands(DOSPlot='APDOS')
calcs.submit()
\end{python}
  \caption{\small 
    Workflow to compute total energies, tight-binding hamiltonians, band structures, densities of states (DOS), and atom projected DOS with and without ACBN0 for  AlP, GaP, InP, AlAs, GaAs, AlSb, GaSb, and InSb. The construction of this workflow is discussed in the text.}
  \label{wfw1}
\end{figure}

 Specific mechanisms for auto-restart and both manual and automatic error control have been implemented to manage the full set of calculations at once. 
The restart mechanism keeps track of the time a cluster job has been running and prepares calculations so that the electronic structure engine exits cleanly before the wall-time runs out. A wall-time buffer with a default of 10\% of the total wall-time request allows time to save data to disk and, if local scratch option is used, for the temporary files in the local scratch diskspace to be copied to the global scratch. After cleaning up temporary files, the command to resubmit the cluster job is given and then the job exits. There are several locking mechanisms in place to ensure that there is no wasted calculation time.
In addition to auto restarting, if there are major failures of the electronic structure engine, the user can load a calculation set and filter only the jobs where there were errors causing the execution pipeline to stop at a certain point and check what kind of errors occurred, if any. Calculations that failed can be filtered into a subset. Certain actions can be taken to help the subset to complete.

\AFLOWpi's source code includes documentation strings written in the Google Python docstring format for readability. These docstrings are located at the beginning of each function or class. \AFLOWpi\ documentation is generated via the {\tt Sphinx} Python package. {\tt Sphinx} can be used to parse source code written in Python, C, C++ as well as several other programming languages to automatically generate user documentation from docstrings. We use Sphinx to generate \LaTeX\ documentation for \AFLOWpi\ as well as a searchable html documentation.

\section{Generation of tight-binding hamiltonians}

Theory and applications of the methodology used by \AFLOWpi\ to generate tight-binding  (TB) hamiltonians for first principles plane-wave pseudopotential calculations have been discussed in Refs. \citep{curtarolo:art86,curtarolo:art108,curtarolo:art111}.
Our implementation does not require any additional input with respect to the electronic structure calculations and provides the real space representation of the TB matrix in self-contained XML format. The sparse TB matrix is represented on a grid as $H_{\alpha,\beta}({\bf r})$ and can be easily Fourier transformed into 

\begin{equation}
{H}_{\alpha,\beta}({\bf k}) = 
\sum_{{\bf r}}  \exp \left(i{\bf k}\cdot {\bf r}\right) {H_{\alpha,\beta}}\left({\bf r}\right),\nonumber
\end{equation}

and diagonalized to determine the full energy dispersions, $E_n({\bf k})$, with the desired level of resolution in reciprocal space or to compute additional properties associated with derivatives in reciprocal space (linear momenta, and effective masses). The real space representation can also be used directly. We have recently applied these methodologies to study diffusive and ballistic transport as well as the optical properties in the independent particles approximation \cite{pino}.

\section{ACBN0 calculations}

The TB representation  of the electronic structure can be exploited to efficiently compute two-electron integrals for the development of local 
exchange functionals.  \AFLOWpi\ implements the direct and self-consistent evaluation of the on-site
Hubbard $U$ correction that greatly improves the accuracy of standard DFT calculations \cite{curtarolo:art93}.
Due to the accurate TB representation,  the evaluation of  the $U$ parameters for atoms in different chemical environments or close to topological defects becomes trivial and \AFLOWpi\ facilitates the investigation of such systems.
ACBN0 only demands computational resources comparable to regular LDA or PBE calculations. 
We have extensively investigated the improvements and the limits of ACBN0 calculation in Refs. \citep{curtarolo:art103,PhysRevB.93.085135}. Overall, ACBN0 delivers better accuracy for lattice parameters and bulk moduli, improves the energy band gap on semiconductors and insulators as well as the relative position of occupied $d$-manifolds in transition metal oxides, and optimizes the phonon dispersions and associated thermal transport properties.

\begin{figure}[h!]
  \centering
    %\framebox[\columnwidth]{\begin{minipage}{\columnwidth}\vspace{3cm}\end{minipage}}
   \includegraphics[width=\columnwidth]{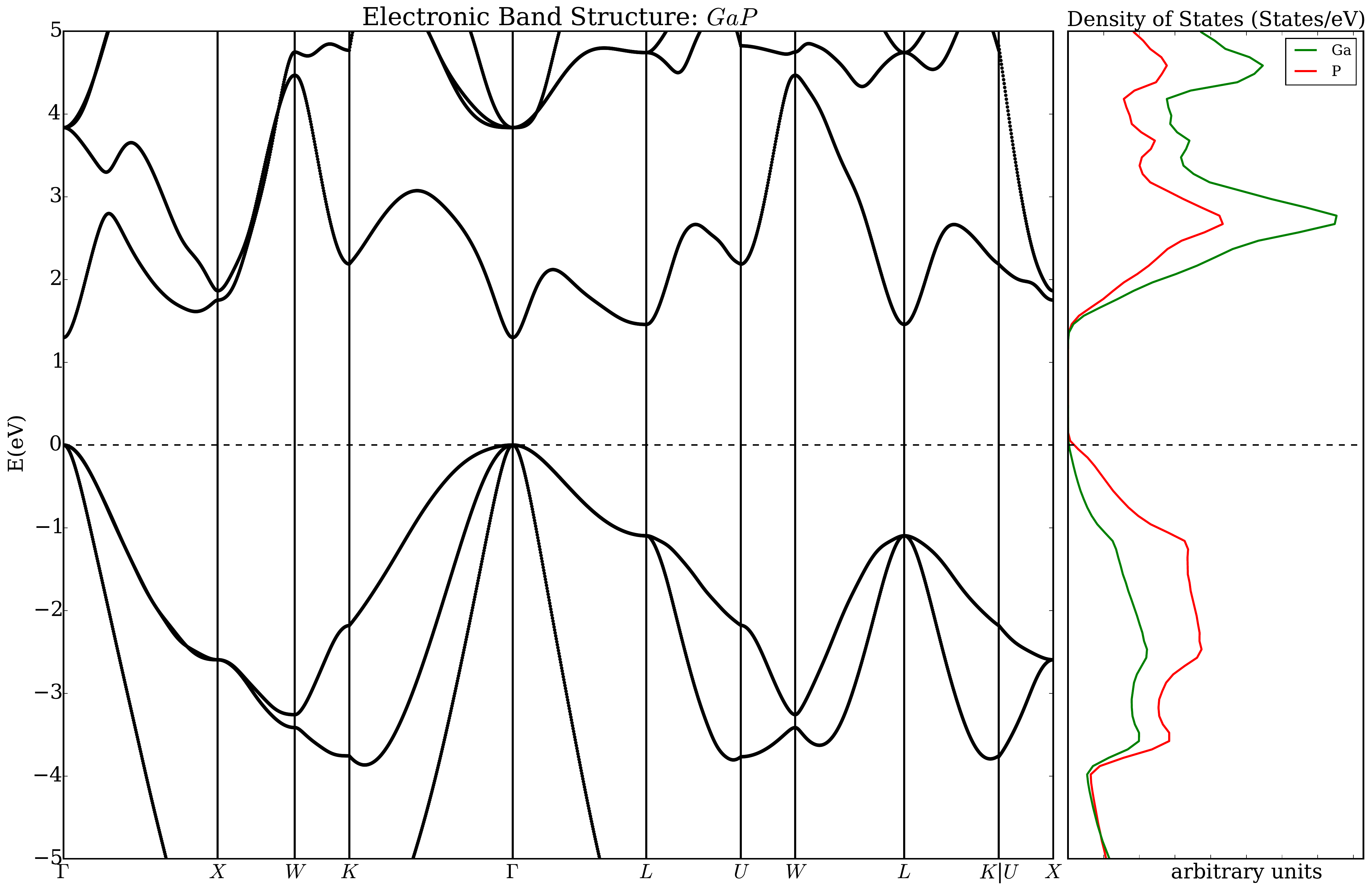}
  \caption{\small 
    Band structure and DOS plot generated by \AFLOWpi\ using the tight-binding (TB) hamiltonian for GaP without ACBN0. The workflow in Figure \ref{wfw1} has been used to obtain this result. TB hamiltonians can be post-processed to perform investigations of the energy dispersions away from symmetry lines or interpolations in reciprocal space at minimal computational cost.}
  \label{bnd1}
\end{figure}
\begin{figure}[h!]
  \centering
  %\framebox[\columnwidth]{\begin{minipage}{\columnwidth}\vspace{3cm}\end{minipage}}
 \includegraphics[width=\columnwidth]{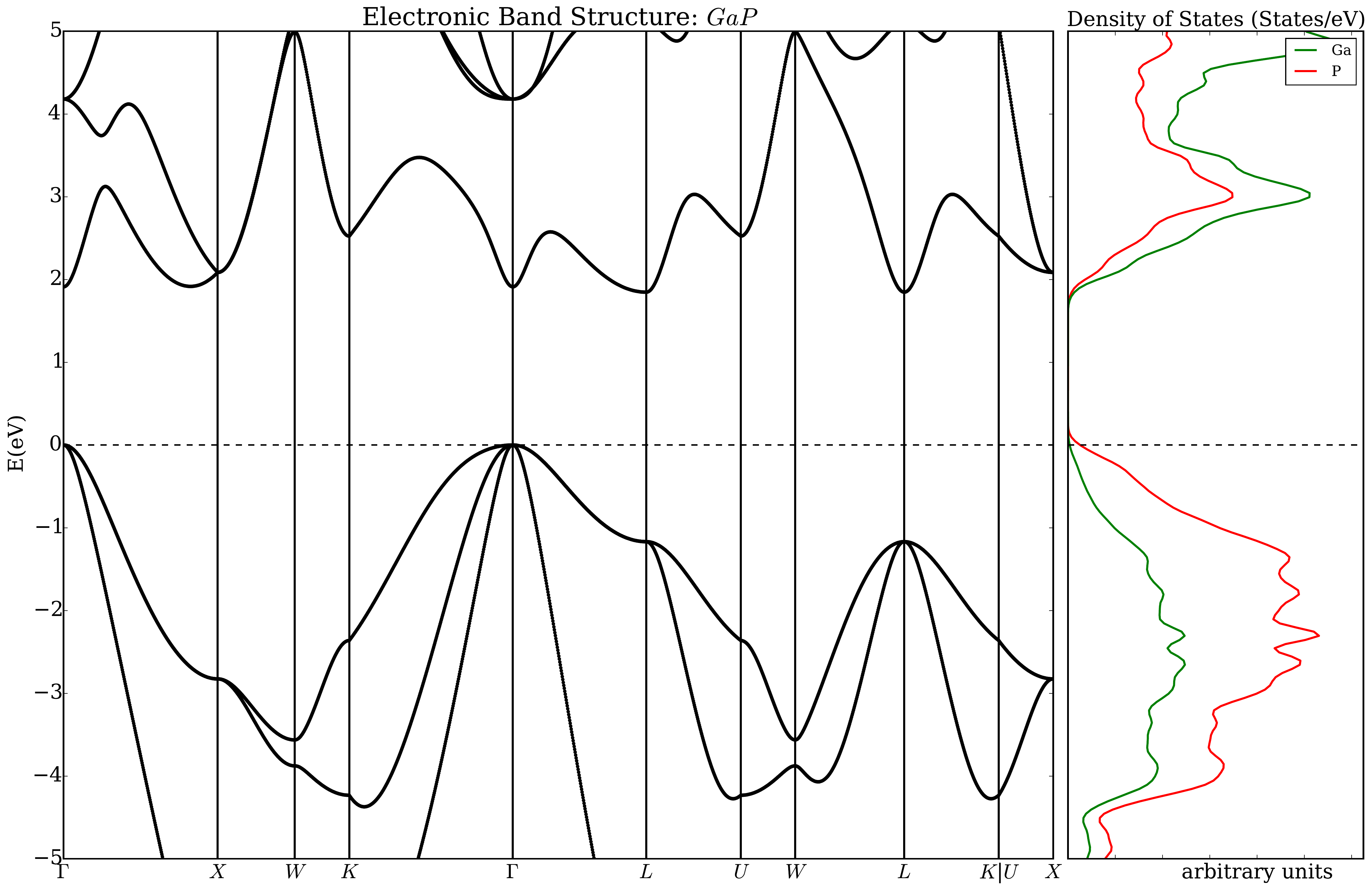}
  \caption{\small 
    As in Figure \ref{bnd1} but computed with ACBN0. The main improvements for this specific example are associated with a more realistic energy gap, and changes in the dispersion curves near $X$ and $L$ in the conduction band. The top valence bands are also slightly more dispersed.}
  \label{bnd2}
\end{figure}

\begin{table*}
  \centering
  \caption{\small 
    Hubbard $U$ corrections (in eV) computed self-consistently for III-V semiconductors with the workflow discussed in the text. The theoretical lattice parameters $ a^{0}_L$  and $ a^{U}_L$ (in \AA) are computed with and without ACBN0 respectively. Energy gaps (in eV) for $\Gamma - \Gamma$, $\Gamma - X$ and $\Gamma - L$ are calculated with and without ACBN0. Experimental values are shown in parentheses.}
  \label{tab-u}
 \bigskip
%  \resizebox{\columnwidth}{!}{%
{\small
    \begin{tabular}{c c c c c c c}
      \hline
       		& AlP & AlAs & AlSb &  GaP & GaAs  & InP \\
\hline
$U_{III}$    & 0.01 & 0.08 & 0.17 & 20.00 & 20.01 & 14.97  \\ 
$U_V $          & 2.31 & 2.12 & 1.11 & 1.87 & 1.75  & 0.00    \\
$ a^{0}_L$ &5.49 (5.46) & 5.74 (5.66)& 6.23 (6.13) & 5.53 (5.45) & 5.79 (6.05) &  5.97 (5.86)  \\	
$ a^{U}_L $ & 5.44 (5.46)& 5.67(5.66) & 6.17(6.13) & 5.48(5.45) & 5.73(6.05) & 5.91 (5.86)  \\   
$ E^{\Gamma}_0 $ & 3.1 (3.6) & 1.8 (3.0) & 1.2 (2.3) & 1.3 (2.8) & 0.0 (0.8 ) & 0.4 (1.4)  \\
$ E^{\Gamma}_U $ & 3.9 (3.6) & 2.5 (3.0) & 1.7 (2.3) & 1.9 (2.8) & 0.5 (0.8) & 1.0 (1.4)  \\ 
$ E^{X}_0$ & 1.3 (2.5) & 1.5 (2.2) & 1.3 (1.7) & 2.1 (2.3) & 1.6 (1.1) & 1.8 (2.4)  \\
$ E^{X}_U$ & 2.1 (2.5) & 1.9 (2.2) & 1.5 (1.7) & 2.1 (2.3) & 1.9 (1.1) & 2.3 (2.4) \\
$ E^{L}_0$ & 2.7 (2.5) & 1.7 (2.4) & 1.3 (2.3) & 1.4 (2.7) & 0.7 (0.9) & 1.1 (2.0)  \\
$ E^{L}_U$ & 3.3 (2.5) & 2.6 (2.4) & 1.5 (2.3) & 1.8 (2.7) & 1.1 (0.9) & 1.8 (2.0) \\     
\hline
  \end{tabular}
%    }}
  }
\end{table*}
\begin{table*}%[t]
  \centering
  \caption{\small Value of the energy gaps and $U$ corrections (in eV) computed with ACBN0 for $AM$N$_2$ compounds.}
  \label{amn2gap}
 \bigskip
  %\resizebox{\columnwidth}{!}{%
{\small
    \begin{tabular}{c c c c c c c}
      \hline
       		&   SrTiN$_2$ &BaTiN$_2$ & SrZrN$_2$ &  BaZrN$_2$ & SrHfN$_2$ & BaHfN$_2$      \\  \hline
      U(A)      &     0.04    & 3.69      & 0.04      &  4.32      & 0.05      & 4.40 \\
      U(M)      &     0.28    & 0.29      & 0.28      &  0.20      & 0.13      & 0.15 \\
      U(N)      &     3.09    & 2.93      & 3.09      &  2.80      & 3.06      & 2.80 \\
      E$_g$     &     1.25    & 1.02      & 1.85      &  1.58      & 2.21      & 1.87 \\
      \hline
    \end{tabular}
    }
\end{table*}
\begin{table*}%[t]%[htdp]
\centering
\caption{\small Elastic constants computed for a set of alkali halides. We determined the $U$ corrections with ACBN0 and  compared with the experimental data from Ref.\ \citep{PhysRev.161.877} and data from Ref.\  \citep{PhysRevB.29.5849}. Structural parameters and $U$ values are in Table\ \ref{halides}.}
\label{elaconshal}
\bigskip
 %\resizebox{\columnwidth}{!}{%
%{\tiny
{\small
\begin{tabular}{c c c c c c}
 \hline
    & $C_{ij}$ ( GPa)  & ACBN0&  PBE & EXP & Model \\
\hline
LiF& $C_{11}$ &113.6 &111.5 &113.9&104.6 \\ 
& $C_{12}$ &50.6&46.3& 47.6 & 42.4 \\ 
& $C_{44}$ &66.2&60.4& 63.6& 70.8\\

LiCl&$C_{11}$  & 51.5 &52.1 &60.7 &66.6   \\
& $C_{12}$& 23.9&21.8& 22.7& 20.9\\
&$C_{44}$ & 27.1&24.9 &26.9&24.6 \\
 
NaF &$C_{11}$ & 101.1 & 95.9&108.9&97.6  \\
&$C_{12}$ & 24.8 &22.2& 22.9& 22.9\\
& $C_{44}$& 29.4&26.4&28.9&29.9  \\

NaCl& $C_{11}$ & 51.2&49.2 &57.3&54.5 \\
& $C_{12}$& 13.4&12.0& 11.2& 11.2\\
& $C_{44}$& 13.9 &12.3&13.3 &11.6\\

KF&   $C_{11}$& 60.9 &60.1& 75.7&85.0\\
& $C_{12}$& 14.9 &14.1 &13.5 & 14.7\\
& $C_{44}$& 14.5 &13.6&13.3&17.1\\

KCl& $C_{11}$ & 37.6 &36.2 &40.9&53.8\\
& $C_{12}$& 6.8 &6.3 &7.0 & 5.4\\
&$C_{44}$ & 7.2 &6.6 &6.2&8.3\\
\hline
\end{tabular}
}
\end{table*}%
\begin{table*}[t]
\centering
  \caption{\small 
    Structural parameters (in \AA) and $U$ corrections (in eV) computed with ACBN0 for selected $AX$ alkali halides using the NaCl structure}
  \label{halides}
 \bigskip
%  \resizebox{\columnwidth}{!}{%
{\small
    \begin{tabular}{cccccccccc}
      \hline
       		&  LiF  &LiCl & LiBr & NaF   & NaCl & NaBr  & KF &  KCl  & KBr \\  \hline
      U(A)      & 0.14 & 0.04 & 0.01 &  0.19 & 0.02 & 0.02  &  0.17  &  0.03 & 0.02 \\
      U(X)      &11.41 & 6.31 & 5.49 & 11.95 & 6.95 & 5.97  & 12.77  &  7.29 & 6.27 \\
     a$_L$      & 3.99 & 5.06 & 5.41 &  4.60 &5.57 &5.89 & 5.32  & 6.24 &6.55 \\ 
     \hline
    \end{tabular}
    }
\end{table*}

\section{Installation and data generation}

The minimalist  style of \AFLOWpi\  makes it easy to install the software. One command to download the source code from its repository online, and two commands to enter the directory and to install \AFLOWpi\ as an importable Python package. The {\tt setup.py} in the base directory of the \AFLOWpi\ source code tree is executed to build and make it importable in Python scripts. All the commands can be run interactively through any Python shell.
Currently \AFLOWpi\ includes three basic submodules: {\tt prep}, {\tt run}, {\tt scfuj}  to  perform the basic calculations including ACBN0 and the generation of tight-binding hamiltonians.  The regular expressions for interacting with a specific electronic structure engine ({\tt pw.x} for instance) are stored separately. In addition, the {\tt plot}, {\tt aflowlib}, {\tt db}, {\tt retr} modules can be used to extract, post-process, and export the computed quantities. The distribution also includes the {\tt pseudo} module for pseudo-potential testing. A detailed description of all the methods is included in the distribution as well as practical examples.
A simple interface class allows for easy customization of calculation workflows using methods to instruct \AFLOWpi\ to prepare the specific tasks.

The construction of a workflow starts  with the command {\tt AFLOWpi.prep.init()} which  initiates the \AFLOWpi\ session and requires a project name and the name of the configuration file as input. Specifying sets within a project is optional but highly recommended. {\tt AFLOWpi.prep.init()} returns a session object which has methods to form the calculation sets. When one of the methods of the session object is called, it returns a ``calculation\_set'' object. The calculation\_set object contains methods to construct workflows which add tasks to be done for each calculation in the set.
Let assume we plan to generate the PAO-TB hamiltonians for the III-V semiconductors (AlP, GaP, InP, AlAs, GaAs, AlSb, GaSb, and InSb) from the electronic structure generated without and with the ACBN0 approach. In a prototypical reference input file ({\it e.g.} {\tt mHT\_TB.ref}) the labels for the atomic species are substituted with the variables {\tt \_AFLOWPI\_A} and {\tt \_AFLOWPI\_B}. That input combined with the list of chemical substitutions is processed by \AFLOWpi\ in order to prepare a pipeline of calculations  starting with the determination of the ground state potential ({\tt session.scfs}), the structural relaxations ({\tt calcs.vcrelax}), and the generation of the TB hamiltonian ({\tt calcs.tight\_binding}) including plots for the densities of states and the band structures. Then the workflow continues with the self-consistent calculation of the Hubbard corrections ({\tt calcs.acbn0}) before recalculating a new TB hamiltonian and new plots.
At the end of the user's  \AFLOWpi\ script, the command {\tt calcs.submit} must be included to run the calculation workflow (see Figure \ref{wfw1}).
In Figs. \ref{bnd1} and \ref{bnd2}, we show plots generated by \AFLOWpi\ before and after the ACBN0 cycle.
It is interesting to pinpoint the corrections of the band gap and the modification of the dispersion relationships (the Hubbard corrections determined running this workflow are in Table \ref{tab-u}). All the tight-binding hamiltonians are stored as XML files and can be distributed independently. We stress the fact that the energy dispersion $E_n({\bf k})$ can be computed with {\it ab initio} accuracy.

Several styles may be used in \AFLOWpi\ to define the calculations in a set.
The first style is based on the cartesian product of two lists (as in Figure \ref{wfw1}). The second style, labelled ``zip mode'', generates all the entries by taking, in order, the corresponding entries in the keywords lists (see Section\ \ref{phonons}). In addition, calculation sets can be formed from a list of one or more input filenames for the electronic structure engine (these can be completed appropriately with the missing variables).

\section{Data retrieval and exporting}

 \AFLOWpi\  extracts and monitors data such as energy, structure, atomic positions, total force, total stress, and automatically records it in the {\tt summary.log} file in the calculation set's {\tt AFLOWpi} directory. Quantities such as those listed above can be extracted for a given step of the calculation pipeline using the {\tt AFLOWpi.retr} module. 

In addition, results can be packaged for the online repository {\tt www.aflow.org} by adding
\bigskip
\begin{python}
AFLOWpi.aflowlib.export('export_example',
                        set_name='III-V',
                        config='./mHT_TB.config')
\end{python}

to the workflow.

\AFLOWpi\  also includes post-processing routines to compute and plot the radial distribution function, compare the energetics of different distortions, generate CIF files,  and perform symmetry analysis.

\section{Examples}
\label{workflows}

\subsection{Optical (and transport) properties}
\label{electronictransport}

Layered nitrides with chemical formula $AM$N$_2$ were studied earlier from the point of view of thermoelectric potentials and superconductivity \cite{C5TA00546A,Henderson1995,doi:10.1021/cm034502y,GREGORY199862,doi:10.1021/cm403840e,doi:10.1021/ic971556z}. As an example we analyze the same chemical space ($A$=Sr, Ba and M=Ti, Zr, Hf) from the point of view of the electronic transport in the constant relaxation time approximation and the optical properties  within the independent electronic approximation.  
Using well converged basis sets and k-space sampling (see Ref.\ \citep{C5TA00546A} for computational details), we have determined the electrical conductivity, the Seebeck coefficient, and  the the complex dielectric functions 
with and without ACBN0. The workflow is in Figure\ \ref{wfw2}.
\begin{figure}[h]
\begin{python}
import AFLOWpi
session = AFLOWpi.prep.init('Transport', 'Si', config='./transport.config')
calcs = session.from_file('transport.in')
calcs.vcrelax()
# get PAO-TB hamiltonian
calcs_TB = calcs.tight_binding()
# optical and transport properties
calcs_TB.epsilon()
calcs_TB.transport(temperature=[300,400])
# plot optical and transport properties
calcs_TB.plot.transport()
calcs_TB.plot.epsilon()
calcs.submit()
\end{python}
  \caption{\small
    Workflow to compute optical (and transport) properties for layered nitrides with chemical formula $AM$N$_2$. The complex dielectric constant is computed within the independent particle approximation with and without the ACBN0 correction The comments show how to compute the electronic transport as discussed in Ref. \citep{pino}.}
  \label{wfw2}
\end{figure}

\begin{figure*}[t]
  \centering
   % \framebox[\textwidth]{\begin{minipage}{\textwidth}\vspace{7cm}\end{minipage}}
\includegraphics[width=\columnwidth]{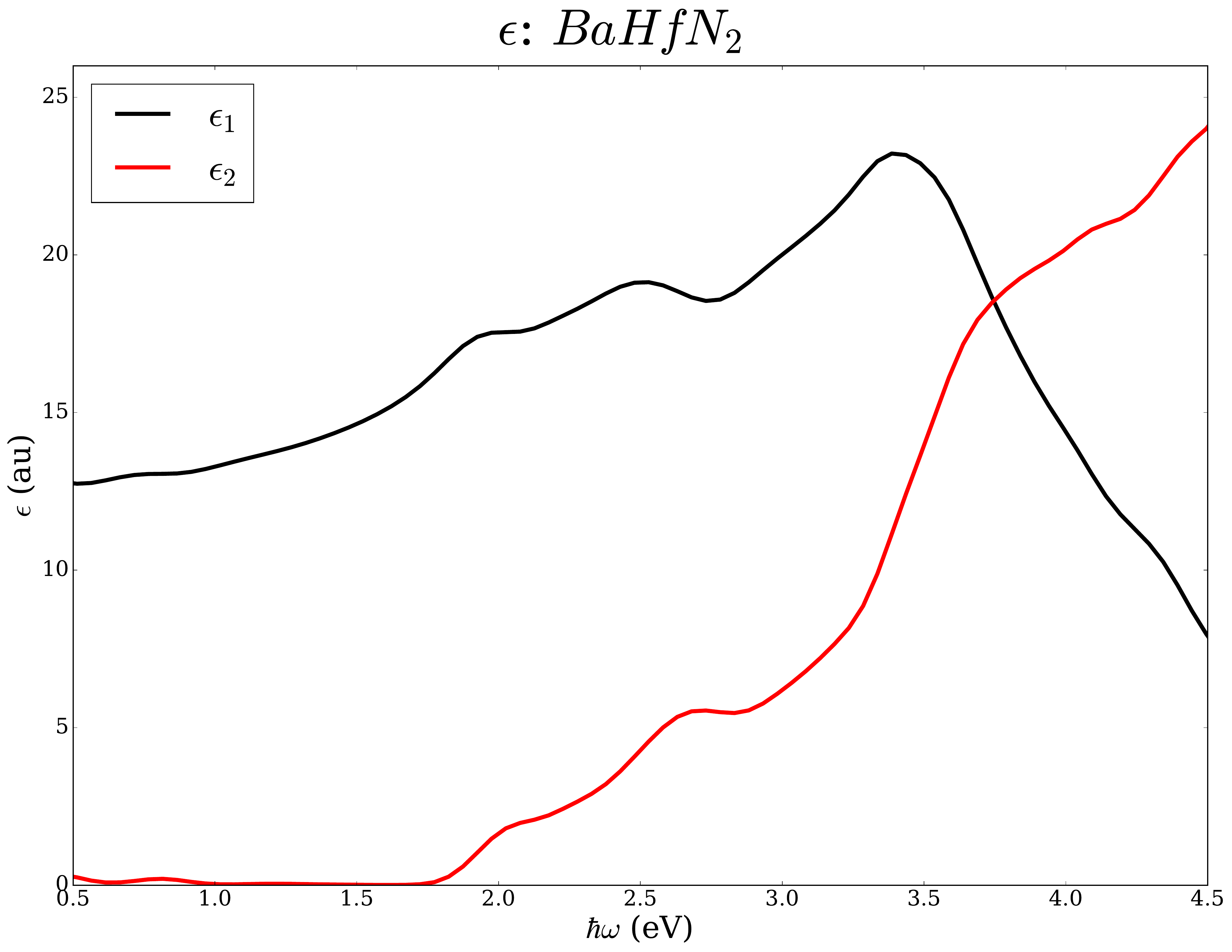}
\includegraphics[width=\columnwidth]{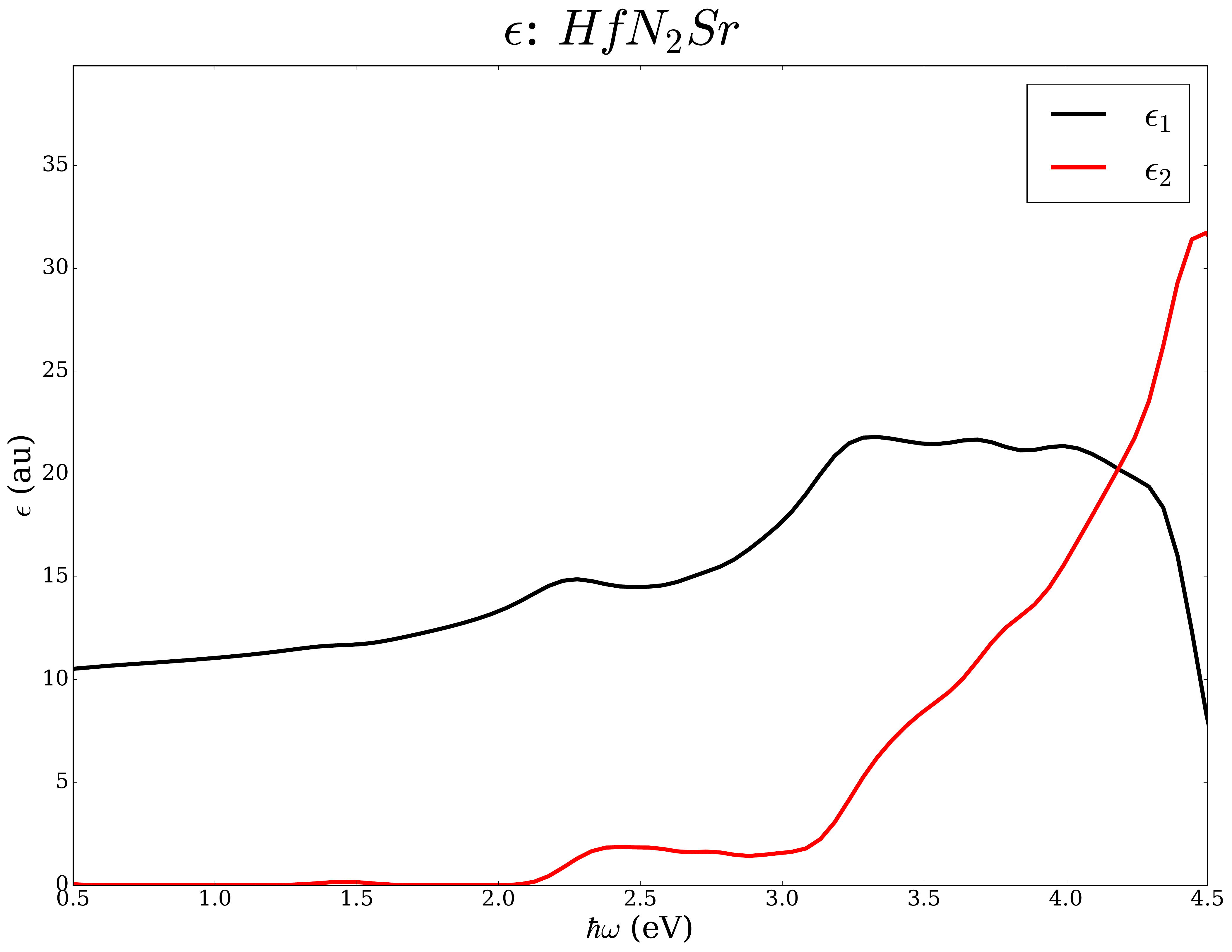}
\includegraphics[width=\columnwidth]{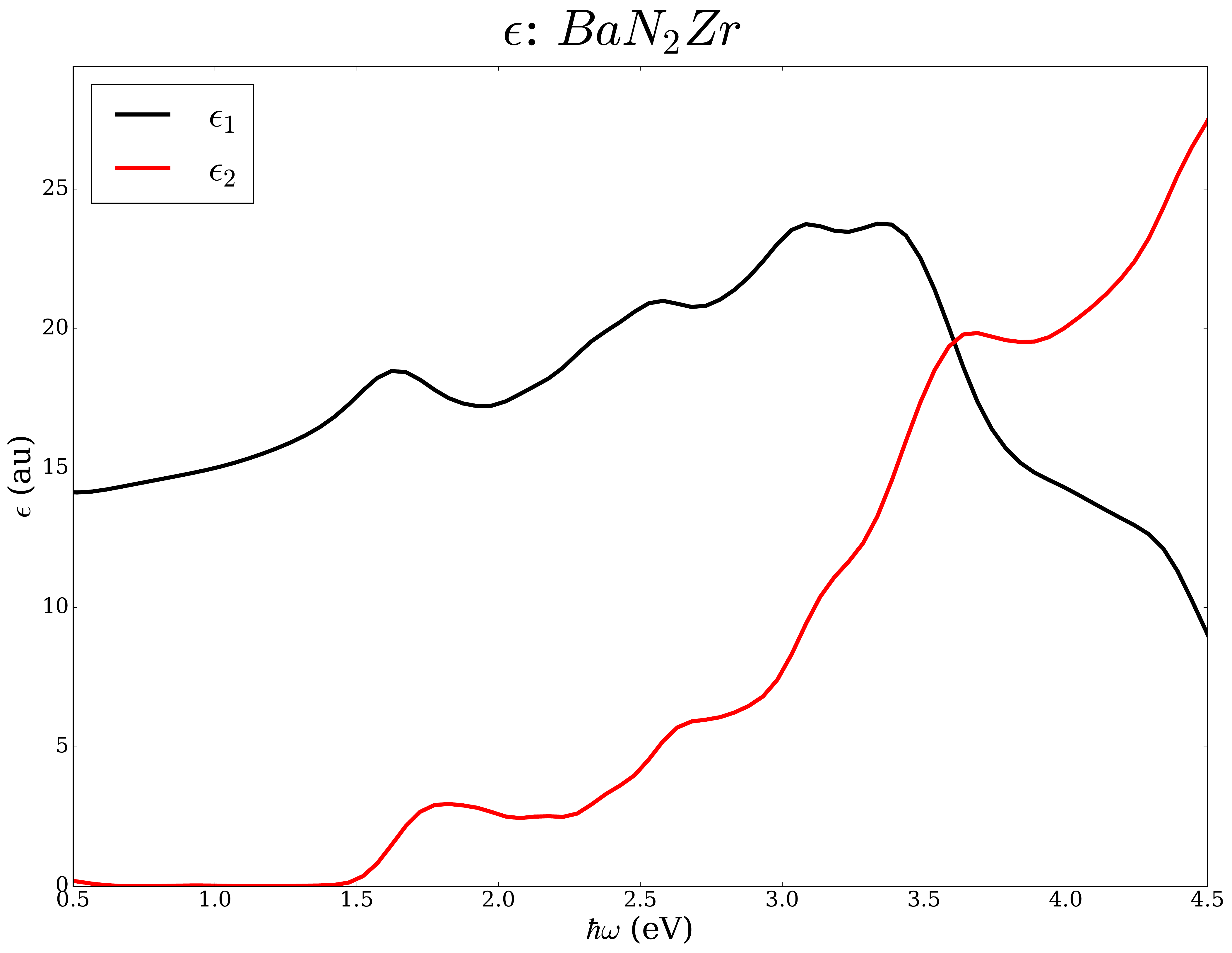}
\includegraphics[width=\columnwidth]{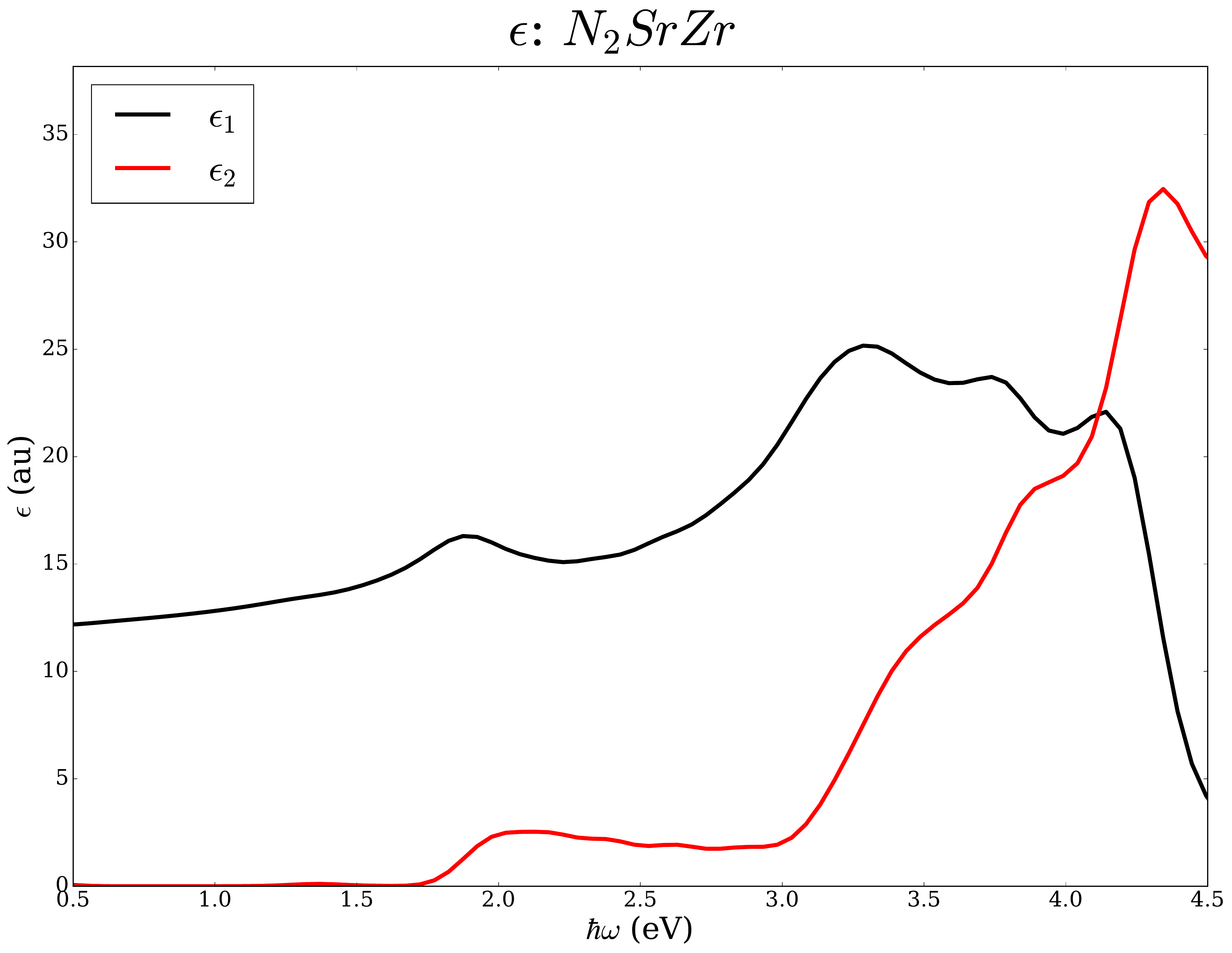}
  \caption{\small 
    Complex dielectric constant generated by \AFLOWpi\ using the tight-binding hamiltonian as in Ref. \citep{pino} for $AM$N$_2$ compounds with $A$=Ba, Sr and $M$=Hf, Zr. 
    The titanates exhibit absorption edges just above 1eV. The workflow in Figure \ref{wfw2} has been used to obtain this result. Additional data including the correction to the band gap provided by ACBN0 are in Table\ \ref{amn2gap}.
}
  \label{eps}
\end{figure*}

We report data for $AM$N$_2$ compounds in Table\ \ref{amn2gap} and in Figure\ \ref{eps}. Among the $AM$N$_2$ compositions, the compounds containing Ti have absorption edges below the Shockley-Queisser limit of 1.34 eV;  
the compounds containing Hf and Zr exceed such a limit but seem potentially interesting as end points of $A$(Ti,$M$)N$_2$ alloys with $M$=Zr and Hf where tuning of the band gap can lead to improved photovoltaics potentials.

\subsection{Phonons and elastic constants}
\label{phonons}

In order to show how the calculations of phonon dispersions and of the elastic constants can be performed in \AFLOWpi\,  we analyzed a small chemical space of alkali halides $AX$ with ($A$=Li, Na, K and $X$=F, Cl, Br) with the NaCl and the CsCl structures. The workflow is in Figure\ \ref{wfw3}: notice the use of the zip mode to specify the chemical space and the use of two reference files for the NaCl and CsCl structure, respectively. Elastic constants are computed solely on the NaCl structure.
\begin{figure}
\begin{python}
import AFLOWpi
# Create AFLOWpi session
ses_CsCl = AFLOWpi.prep.init('Halide', 'CsCl',
                            config='./phonon.config')
ses_NaCl = AFLOWpi.prep.init('Halide', 'NaCl',
                            config='./phonon.config')
# Generate a calculation set from a reference input file
allvars={"\_AFLOWPI\_A_":("Li","Na","K",),
  "\_AFLOWPI\_B_":("F","Cl","Br","I"),}
CsCl = ses_CsCl.scfs(allvars,'CsCl.ref',)
NaCl = ses_NaCl.scfs(allvars,'NaCl.ref',)
#do a vc-relax
CsCl.vcrelax()
NaCl.vcrelax()
# change the thresholds in the input files
# changing input for a step are called after
# the call for the step. the change_calcs
# below affect the vc-relax above
CsCl.change_input("&control","etot_conv_thr","1.0D-5")
CsCl.change_input("&control","forc_conv_thr","1.0D-4")
CsCl.change_input("&electrons","conv_thr","1.0D-16")
NaCl.change_input("&control","etot_conv_thr","1.0D-5")
NaCl.change_input("&control","forc_conv_thr","1.0D-4")
NaCl.change_input("&electrons","conv_thr","1.0D-16")
#do another relax just to be safe
CsCl.vcrelax()
NaCl.vcrelax()
#do phonon calculations with 2x2x2 supercell
CsCl.phonon(mult_jobs=True,nrx1=3,nrx2=3,nrx3=3,innx=2, de=0.003,LOTO=True,field_strength=0.003, disp_sym=True,atom_sym=False,)
NaCl.phonon(mult_jobs=True,nrx1=3,nrx2=3,nrx3=3,innx=2, de=0.003,LOTO=True,field_strength=0.003, disp_sym=True,atom_sym=False,)
#plot the phonons
NaCl.plot.phonon(postfix='NaCl')
CsCl.plot.phonon(postfix='CsCl')
# do the elastic constants with the ElaStic Package 
# Install: http://exciting-code.org/elastic
NaCl.elastic(mult_jobs=False,num_dist=10,eta_max=0.001)
#do acbn0 to do phonon with self-consistently
#determined Hubbard U. Changing the cell while
#converging the U should be done with caution.
NaCl.scfuj(relax='vc-relax')
CsCl.scfuj(relax='vc-relax')
NaCl.elastic(mult_jobs=False,num_dist=10,eta_max=0.001)
NaCl.phonon(mult_jobs=True,nrx1=3,nrx2=3,nrx3=3,innx=2, de=0.003,LOTO=True,field_strength=0.003, disp_sym=True,atom_sym=False,)
CsCl.phonon(mult_jobs=True,nrx1=3,nrx2=3,nrx3=3,innx=2, de=0.003,LOTO=True,field_strength=0.003, disp_sym=True,atom_sym=False,)
CsCl.plot.phonon(postfix='CsCl_acbn0')
NaCl.plot.phonon(postfix='NaCl_acbn0')
# submit the calcs to run
calcs.submit()
\end{python}
  \caption{\small 
Workflow to compute phonon dispersion with and without ACBN0 for few alkali halides in NaCl and CsCl structure. For the NaCl set of calculations also the elastic constants are computed and compared with the experimental data in Table\ \ref{elaconshal}.}
  \label{wfw3}
\end{figure}

\begin{figure*}
  \centering
%\framebox[\textwidth]{\begin{minipage}{\textwidth}\vspace{10cm}\end{minipage}}
\includegraphics[scale=0.3]{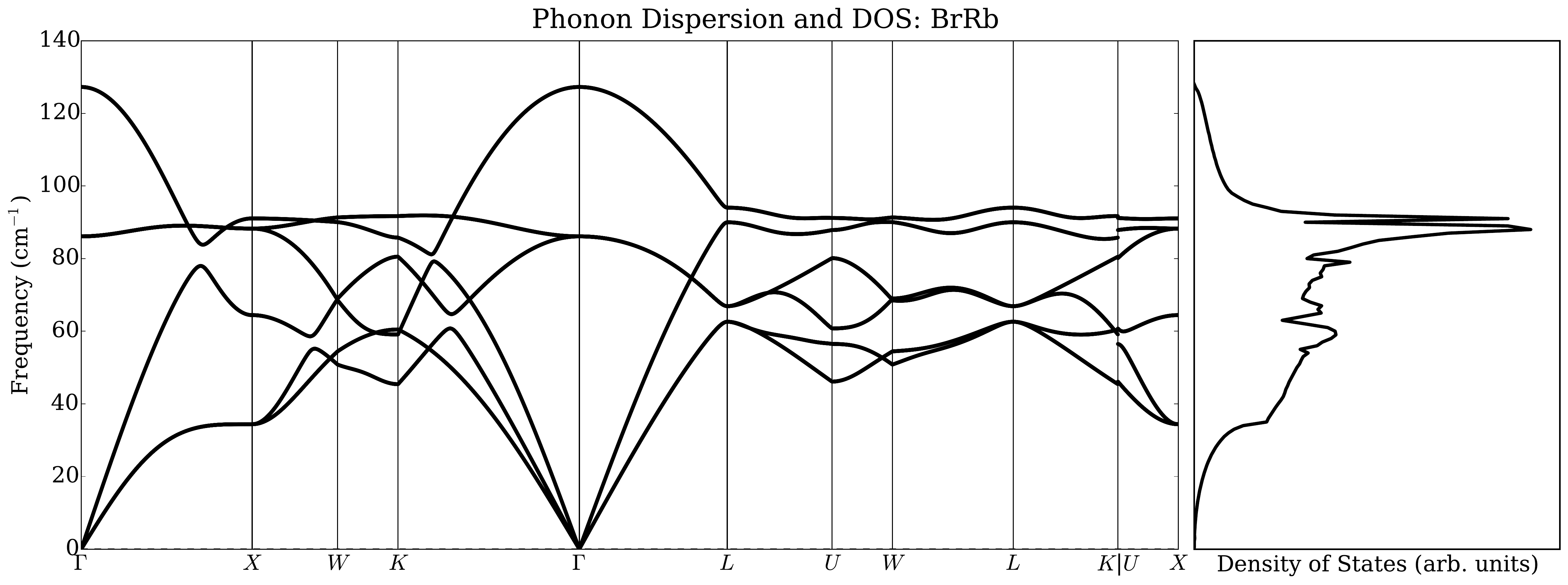}
\includegraphics[scale=0.3]{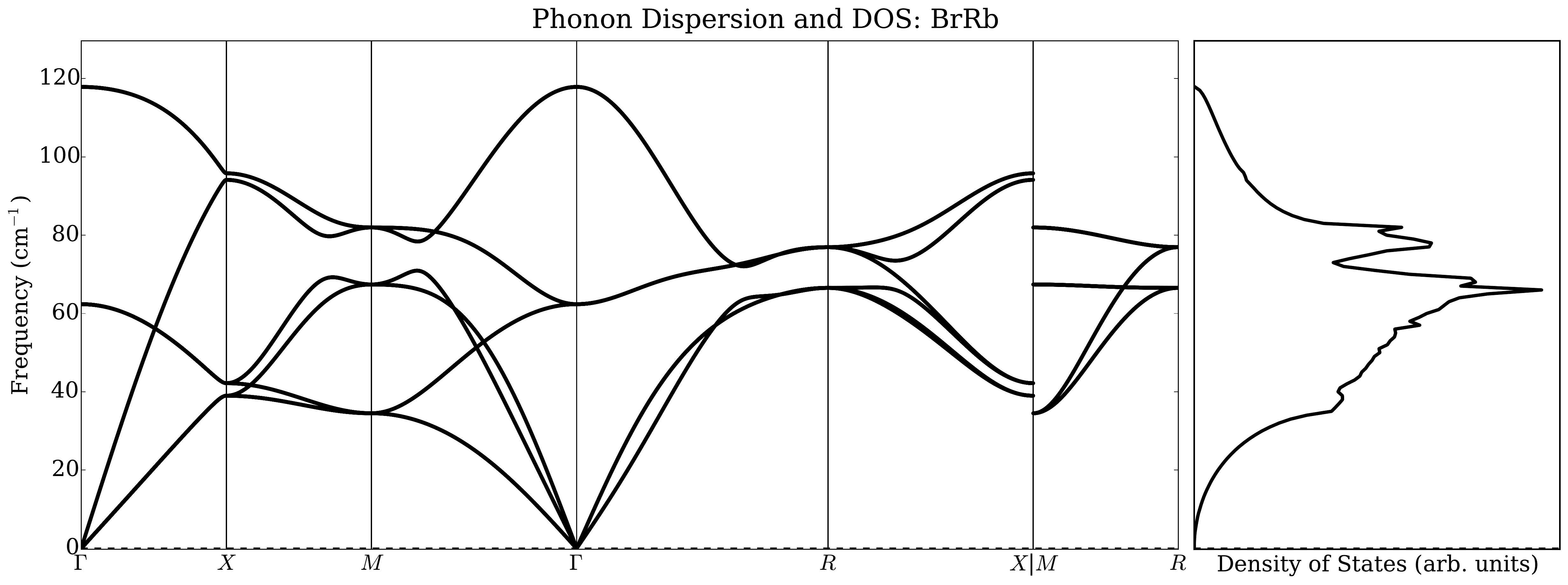}
\includegraphics[scale=0.3]{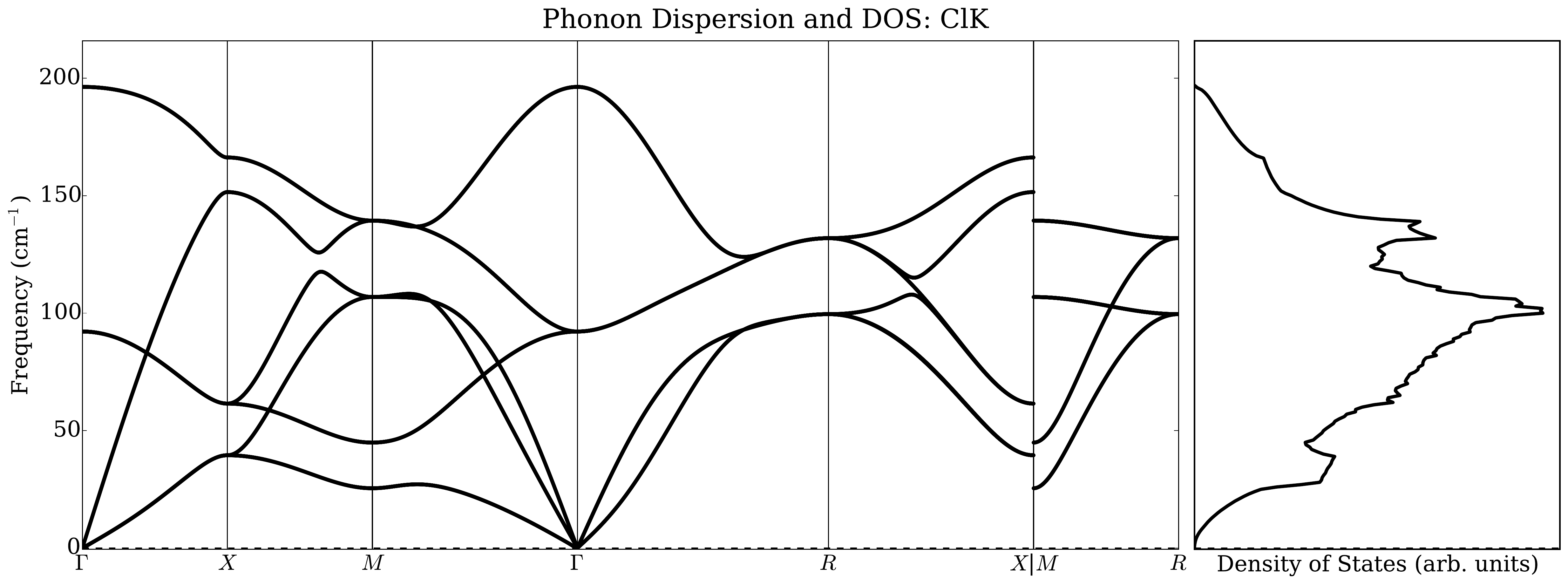}
\includegraphics[scale=0.3]{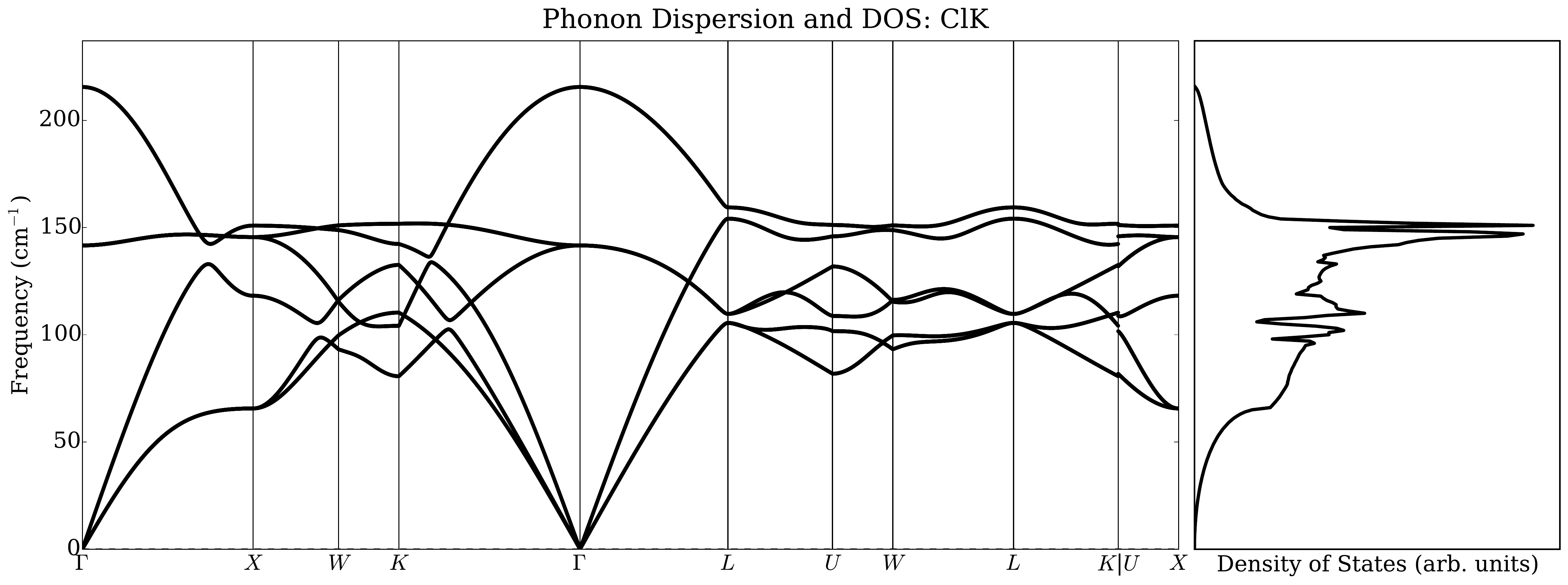}
  \caption{\small 
    Phonon branches and vibrational density of states for CsCl and NaCl structure of RbBr and KCl at equilibrium lattice parameters. We used the workflow in Figure \ref{wfw3}.}
  \label{ph}
\end{figure*}

Under normal conditions, alkali halides have NaCl or B1-type crystal structure, but they transform under moderate pressure to the CsCl or B2-type structure. There has been much work done on thermal conductivity during the phase transition of alkali halides: Averkin \cite{averkin1977thermoconductivity} found that the phase transition completely eliminates the influence of the elastic anisotropy on the thermal conductivity, as well as its dependence on pressure. In that paper an empirical relation was also established between elastic anisotropy and thermal conductivity. However, Slack \cite{0022-3719-18-20-021} concluded that there is no need to invoke the elastic anisotropy to explain thermal conductivity differences. He noticed that NaCl structure compounds have a higher thermal conductivity ratio than CsCl ones by about a factor of two and conjectured a large increase in the Gr\"uneisen parameter as the co-ordination number changes from six to eight. Our results indicate, however, that a major contribution to the change in the thermal conductivity is related to the average sound velocity reduction in CsCl structure with respect to the NaCl structure. In Figure\ \ref{ph} we report, as an example, phonon dispersions automatically generated by \AFLOWpi\ using a finite difference method. We did not observe significant changes using the ACBN0 approach with respect to the PBE calculations, and we choose to report selected ACBN0 data as an example.

The elastic constants of alkali halides are important as they are closely related to the ionic binding force in the crystal \cite{PhysRevB.29.5849}. There are several phenomenological models to determine the elastic constants: in the central force model \cite{Michihiro2010572}, the elastic constants are given by the sum of the terms due to long range and short range central forces. This model leads to a  Cauchy relation $C_{12} = C_{44}$, which is not true experimentally. The degree to which this Cauchy relation is not satisfied is a measure of the importance of many-body forces in lattice dynamics of the alkali halide which Lowdin \cite{PhysRev.97.1490} has found to exist. The breathing shell model \cite{SCHRODER1966347} includes many-body forces and makes a prediction $C_{12} < C_{44}$ in alkali halides. Mahan \cite{PhysRevB.29.5849} studied the effect of induced polarization on phonon modes at long wavelength and introduced a polarization term in the central force model predicting $C_{12}>C_{44}$. Experimentally the deviation is found in both directions. We report in Table \ref{elaconshal} the comparison between first principles calculations with and without ACBN0, experimental data, and results from Ref.\ \citep{PhysRevB.29.5849}. Our results improve the agreement with experimental data with respect to the model in Ref. \ \citep{PhysRevB.29.5849} and indicate the sensitivity of the elastic constants to the choice of the computational method.

\section{Conclusions}
We have presented here a minimalist high-throughput software infrastructure that, in addition to organizing and controlling the computational workflow, facilitates the generation of tight-binding Hamiltonians to represent accurately the electronic properties with \textit{ab initio} accuracy.
\AFLOWpi\ exploits the tight-binding formalism to compute electronic transport and optical properties within the independent particle approximation.
Three examples were used to illustrate the features of \AFLOWpi\: (1) the calculation of the band structure of III-V semiconductors with and without the ACBN0 correction, (2) the study of the complex dielectric constant of $AM$N$_2$ ($A$=Sr, Ba and $M$=Ti, Zr, Hf), and (3) the phonon dispersions and the elastic constants of binary halides 
($AX$ with $A$=Li, Na, K and $X$=F, Cl, Br). We found that $A$(Ti,$M$)N$_2$ alloys with $M$=Zr and Hf may have potential for photovoltaic applications.
In addition, we justified the dramatic changes in thermal conductivity occurring at the transition between NaCl and CsCl structure in terms of sound velocity.
 The code, including the examples presented in this paper, is available at {\tt www.aflow.org/aflowpi}.

\section{Acknowledgments}
{\small
We are grateful to  the High Performance Computing Center at Michigan State University and the Texas Advanced Computing Center at the University of Texas Austin.
The members of the \AFLOW\ Consortium  (http://www.aflow.org)
acknowledge support  by DOD-ONR (N00014-13-1-0635, N00014-11-1-0136,
N00014-15-1-2863). The authors also acknowledge Duke University ---
Center for Materials Genomics --- and the CRAY corporation for
computational support.}

%\section*{References}

\end{document}